\documentclass{article}
\usepackage{spconf,amsmath,amssymb,amsthm,graphicx}
\usepackage{cite}
\usepackage{booktabs}
\usepackage{algorithm}
\usepackage{algpseudocode}
\algrenewcommand\algorithmicrequire{\textbf{Input:}}
\usepackage{xcolor}
\usepackage{url}
\usepackage{subcaption}
\usepackage{hyperref}
\input{mysymbol.sty}

\newcommand{\supp}{\operatorname{supp}}

\renewcommand{\argmin}{\operatorname*{arg\,min}}
\newtheorem{proposition}{Proposition}
\theoremstyle{remark}
\newtheorem{remark}{Remark}
\title{Sparse Experimental Design for Nonsmooth Estimators\\ via Bilevel Optimization}

\name{Antonio G. Marques and Samuel Rey\thanks{Work supported by the Spanish AEI (AEI/10.13039/501100011033) grants PID2022-136887NB-I00 and PID2025-170000NB-I00, and the Community of Madrid via IDEA-CM (TEC-2024/COM-89), L\'INEA A. CP2301 (URJC-F1180), and the Ellis Madrid Unit. Claude AI was used to assist in writing the manuscript and coding the simulations. The authors take full responsibility for the results of this paper.} }

\address{Dept.\ of Signal Theory and Communications, King Juan Carlos University, Madrid, Spain}

\begin{document}
\ninept
\makeatletter
\g@addto@macro\small{\abovedisplayskip 4pt plus 2pt minus 2pt \belowdisplayskip 4pt plus 2pt minus 2pt \abovedisplayshortskip 0pt plus 2pt \belowdisplayshortskip 2pt plus 2pt}
\makeatother
\normalsize
\setlength{\textfloatsep}{7pt plus 2pt minus 2pt}
\setlength{\floatsep}{6pt plus 2pt minus 2pt}

\maketitle

\begin{abstract}
Optimal experimental design (OED) decides which measurements to acquire for downstream estimation. Classically, this is done by optimizing an information criterion derived from a linear-Gaussian model, for which the estimator is available in closed form. Many modern estimators such as Lasso, elastic net, or total-variation-based estimators, however, are \emph{nonsmooth} and lack a closed-form solution. To extend OED to these scenarios, we recast the problem as what it implicitly is: a \textit{bilevel program} whose lower level computes the deployed nonsmooth estimator and whose upper level scores its validation prediction risk. Rather than fixing the number of measurements in advance, we charge each continuous acquisition weight a concave sparsity price. As a result, \emph{how many} and \emph{which} measurements to keep are both outcomes of the optimization. 
A measurement survives only if its estimator-aware value exceeds its price, and progressively increasing the price yields a sequence of designs that explores the trade-off between measurement cardinality and estimation accuracy.
Using a value-function penalty reformulation, we develop a single-loop proximal-gradient algorithm that avoids differentiating the nonsmooth solution map and establish its convergence to an (approximate) stationary point.
Experiments on synthetic sparse recovery and image reconstruction demonstrate the benefits of the proposed method.
\end{abstract}

\begin{keywords}
Optimal experimental design, bilevel optimization, nonsmooth optimization, sensor selection, sparse regularization
\end{keywords}

\section{Introduction}\label{sec:intro}
Deciding which sensors to activate, which samples to keep, or which rows of a measurement operator to acquire is a classical problem in signal processing and statistics, known as optimal experimental design (OED) \cite{pukelsheim2006optimal,kay1993estimation}.
The textbook recipe starts from a linear-Gaussian model and evaluates each candidate design through its Fisher information matrix or posterior covariance. The design is then optimized according to an A-, D-, or E-optimality criterion using greedy methods, submodular optimization, or convex relaxations \cite{joshi2009sensor,krause2008nearoptimal,chepuri2015sparsity,manohar2018datadriven}.
This classical strategy is elegant, scalable, and deeply rooted, but it scores a \emph{model}, not an \emph{estimator}.
In contemporary pipelines, however, the estimator used to process the acquired data is rarely the closed-form Gaussian estimator.
Often, it is defined by a nonsmooth optimization problem, as in Lasso, elastic net, or total-variation (TV) regularized reconstruction~\cite{tibshirani1996lasso,rudin1992nonlinear,combettes2011proximal}.
The performance of these estimators depends on their active-set behavior, shrinkage bias, and support instabilities, effects that the information matrix of the forward model does not explicitly capture.

Our starting point is the observation that OED is, by nature, a \emph{bilevel} optimization (BLO) problem \cite{bard1998practical,colson2007overview}: the upper level (UL) chooses the design (the samples), and the lower level (LL) computes the estimate from the data that design produces.
In the linear-Gaussian setting, the LL solution admits a closed-form expression that can be substituted into the UL objective, yielding a problem in the design variables alone. The bilevel structure then collapses and is easy to overlook.
In contrast, the collapse generally fails when the deployed estimator is nonsmooth, and the design problem becomes a genuine BLO problem with a nonsmooth LL.
Mainstream BLO methods \cite{maclaurin2015gradient,franceschi2018bilevel,liu2021valuefunction,shen2023penalty,lu2024firstorder,zhang2024introduction} are mostly built on smooth or strongly regular LL problems and do not directly apply to this setting.
Related bilevel formulations have emerged in imaging and inverse problems, including regularizer learning, sampling-pattern optimization, and constrained-design formulations \cite{kunisch2013bilevel,ruthotto2018optimal,sherry2020learning,ehrhardt2021inexact,crockett2022bilevel}. However, they typically smooth the LL, differentiate through a truncated solver, or rely on support identification \cite{bertrand2022implicit}.

This paper develops a BLO-based formulation for estimator-aware OED with three distinguishing features:
a) the LL is the \emph{exact} nonsmooth estimator that will be deployed, with no smoothing or unrolling involved; b) the UL is a validation prediction risk, computable without ground-truth signals and meaningful for any forward model;
and c) instead of fixing a cardinality budget upfront, we constrain the acquisition weights to the box $[0,1]^M$ and assign them a concave sparsity price.
Under c), measurements must \emph{pay for themselves} and the optimization decides how many survive.
Moreover, progressively increasing the price yields a sequence of designs that explores the trade-off between measurement cardinality and estimation accuracy.
Algorithmically, we adapt the value-function penalty methodology for \textit{nonsmooth} BLO of \cite{rey2026singlelevel}, where the bilevel optimality constraint is replaced by a penalized optimality gap, yielding a single-loop proximal-gradient scheme.
The scheme requires only inexact LL solves and computationally inexpensive proximal steps, avoids differentiating the LL solution map (so that hypergradients are not needed), and inherits an $O(1/K)$ stationarity guarantee.
Its stationarity conditions expose an interpretable keep-or-discard rule: a sensor is retained only when the improvement in the fit of its measurement demanded by the UL exceeds its marginal price.
Experiments on synthetic sparse recovery and Fashion-MNIST pixel selection evaluate the proposed approach against classical information-based and heuristic selection methods at matched cardinality.

\section{OED is a bilevel problem}
\label{sec:oed_blo}
Let $\bbw\in[0,1]^M$ denote the acquisition weights for $M$ candidate measurements, with $w_i=0$ indicating that measurement $i$ is not acquired, and let $\bbbeta\in\reals^D$ denote the coefficients to be estimated.
For a given design $\bbw$, any approach that scores $\bbw$ by the quality of the resulting estimate is implicitly solving the following BLO problem
\begin{equation}\label{eq:blo}
	\min_{\bbw\in[0,1]^M}\ \Phi\bigl(\bbw,\bbbeta^\star(\bbw)\bigr)
	\ \ \text{s.t.}\ \
	\bbbeta^\star(\bbw)\in\argmin_{\bbbeta}\ g(\bbw,\bbbeta),
\end{equation}
where the LL cost $g$ defines the estimator and the UL cost $\Phi$ quantifies its performance under the selected design $\bbw$. Classical OED never writes \eqref{eq:blo} because two simplifications remove the LL. Consider a linear model $\bby=\bbX\bbbeta+\bbepsilon$, with $\bbepsilon\sim\ccalN(\bbzero,\bbR)$ and diagonal covariance $\bbR \in \reals^{M \times M}$.
Under a Gaussian prior, the LL solution is the weighted ridge estimator, which is available in closed form and can be \emph{analytically eliminated}.
Furthermore, the classical A-, D-, and E-optimality criteria are defined by the functionals
\begin{equation}\label{eq:ade}
\Phi_D=-\log\det\bbF,\quad \Phi_A=\tr(\bbF^{-1}),\quad \Phi_E=-\lambda_{\min}(\bbF)
\end{equation}
of the information matrix $\bbF(\bbw)=\bbX^\top\diag(\bbw)\bbR^{-1}\bbX+\mu\bbI$. $\bbF(\bbw)$ depends on the design but \emph{not on the estimator}, so that the UL and LL decouple. As a result, \eqref{eq:blo} collapses to a single-level program in $\bbw$, which is the form in which OED is usually formulated \cite{pukelsheim2006optimal,joshi2009sensor}.

Both simplifications fail for the nonsmooth estimators analyzed in this paper.
To ease exposition, we focus on the design-weighted elastic net estimator for the LL.
Formally, assume that we have $n=1,\ldots,N$ signal instances (an ensemble of training scenes, channel snapshots, or images) that share the design $\bbw$. The optimal estimator for each of those instances (say the $n$-th one) is found as
\begin{subequations}\label{eq:ll}
\begin{align}\label{eq:ll_objective}
\bbbeta^\star_n(\bbw)=&\argmin_{\bbbeta}\
\underbrace{\sum_{i=1}^M \tfrac{w_i}{R_{ii}}\,\ell_{n,i}(\bbbeta)+\tfrac{\mu}{2}\|\bbbeta\|_2^2}_{g_{{\rm s},n}(\bbw,\bbbeta)}
+\underbrace{\lambda\|\bbbeta\|_1}_{g_{{\rm n},n}(\bbbeta)},\\
&\hspace{1.3cm}\text{with}\;\;\ell_{n,i}(\bbbeta)=\tfrac12(y_{n,i}-\bbx_i^\top\bbbeta)^2,\label{eq:ll_objective_specified_linReg}
\end{align}
\end{subequations}
where $\bby_n=\bbX\bbbeta_n+\bbepsilon_n$ collects the $M$ candidate measurements of instance $n$ and $\bbx_i^\top$ is the $i$-th row of $\bbX$. No closed form exists for the generic LL problem in \eqref{eq:ll}. In addition, the map $\bbw\mapsto\bbbeta_n^\star(\bbw)$ is nonsmooth, so \eqref{eq:blo} must be faced as a bilevel program.
The algorithmic development relies on two structural properties of \eqref{eq:ll}. First, the design (UL variable) enters only the \emph{smooth} component $g_{{\rm s},n}$, which is $\mu$-strongly convex in $\bbbeta$ uniformly over $\bbw\ge\bbzero$.
Second, the nonsmooth term $g_{{\rm n},n}$ is design-independent (it depends only on the LL variables) and proximable.
Furthermore, while \eqref{eq:ll_objective_specified_linReg} specializes to a sparse linear regression, the proposed framework only requires the LL cost to have the form $g_n(\bbw,\bbbeta)=\sum_i \tfrac{w_i}{R_{ii}}\ell_{n,i}(\bbbeta)+h(\bbbeta)+g_{{\rm n},n}(\bbbeta)$ with: a) $\ell_{n,i}$ convex and smooth in $\bbbeta$ (e.g., squared, logistic, or Huber losses); b) $h$ smooth and $\mu$-strongly convex; and c) $g_{{\rm n},n}$ convex, design-independent, and proximable.
These assumptions ensure that the design enters only through the weights of the smooth per-measurement losses, while strong convexity stems from $h$ and hence holds for every $\bbw$ in the box.
The template covers Lasso ($\mu\to0^+$, kept positive for strong convexity), transform-domain sparsity and, replacing the proximal operator, TV and fused-Lasso regularizers \cite{tibshirani2005fused,combettes2011proximal}.

\section{Sparsity-promoting estimator-aware design}\label{sec:formulation}
\subsection{Upper level: validation risk plus a sparsity price}
Since ground-truth signals are unavailable in deployment, we score a design by the \emph{prediction risk} of its reconstructions on held-out measurements.
Specifically, given validation pairs $\{(\bbX_{{\rm v},n},\bby_{{\rm v},n})\}_{n=1}^N$ for the same instances, we define the empirical prediction risk as
\begin{equation}\label{eq:phi}
\Phi\bigl(\bbbeta_1,\ldots,\bbbeta_N\bigr)=\frac{1}{N}\sum_{n=1}^N \tfrac12\bigl\|\bbX_{{\rm v},n}\bbbeta_n-\bby_{{\rm v},n}\bigr\|_2^2 ,
\end{equation}
evaluated at $\bbbeta_n=\bbbeta_n^\star(\bbw)$.
This criterion is estimator-aware by construction since it can only be lowered by designs $\bbw$ that improve the accuracy of the \emph{deployed} estimator. Moreover, it needs neither Gaussianity nor a Jacobian of the forward model.

The second ingredient is how sparsity of the design $\bbw$ arises. A common approach fixes a cardinality $M_1$, optimizes over a relaxed budget set, and rounds the solution~\cite{joshi2009sensor,chepuri2015sparsity}. We take the complementary, budget-free route, where acquisition is charged a separable sparsity price.
This yields the design problem
\begin{equation}\label{eq:sparse_blo}
	\min_{\bbw\in[0,1]^M}\ \Phi\bigl(\bbbeta_1^\star(\bbw),\ldots,\bbbeta_N^\star(\bbw)\bigr)+\eta\sum_{i=1}^M\varphi(w_i),
\end{equation}
with $\bbbeta_n^\star(\bbw)$ as in \eqref{eq:ll}, a price level $\eta>0$, and an increasing, concave function $\varphi:[0,1]\to[0,1]$ satisfying $\varphi(0)=0$.
While the linear choice $\varphi(w)=w$ gives the $\ell_1$ price $\eta\|\bbw\|_1$, here we advocate the transformed $\ell_1$ device for sharpening sparsity~\cite{zhang2018transformed}, given by the concave function
\begin{equation}\label{eq:concave_price}
	\varphi_\theta(w)={w}/{(w+\theta)},\qquad \theta>0.
\end{equation}
The distinction matters here for a structural reason. Under the linear price, the gradient [see \eqref{eq:wgrad} for the detailed expression] pushes every weight down by the same amount $\eta$, regardless of its current value. Lowering all weights uniformly is almost equivalent to increasing the regularization level in \eqref{eq:ll}, so the design first degrades all reconstructions and only later starts switching sensors off. The concave price avoids this: its derivative $\varphi_\theta'(w)=\theta/(w+\theta)^2$ is small for $w\approx1$ and large near $0$, so weights close to 1 are barely penalized while small weights are driven to exactly zero.
Another option for promoting sparsity is reweighted \(\ell_1\) minimization~\cite{candes2008enhancing,buciulea2022learning}.
Because the projection onto the box produces exact zeros, the deployed selection is the support of $\bbw$. The number of active measurements is thus an output, and the price path $\{\bbw^\star(\eta)\}$ sweeps the frontier between acquisition cost and reconstruction performance.

\subsection{Value-function penalty reformulation}
Problem \eqref{eq:sparse_blo} is a nonsmooth BLO problem, and the standard machinery (implicit differentiation, unrolling) stumbles on the nondifferentiable solution map \cite{bertrand2022implicit}. Following \cite{rey2026singlelevel,liu2021valuefunction,shen2023penalty}, we instead work with the LL \emph{value functions} $v_n(\bbw)=\min_{\bbbeta} g_n(\bbw,\bbbeta)$, where $g_n(\bbw,\bbbeta)=g_{{\rm s},n}(\bbw,\bbbeta)+g_{{\rm n},n}(\bbbeta)$. Because the design multiplies only the smooth, strongly convex fidelity, each $v_n$ is differentiable with Lipschitz gradient despite the $\ell_1$ term. Moreover, leveraging the results in \cite{rey2026singlelevel}, a Danskin-type reasoning gives the remarkably simple expression
\begin{equation}\label{eq:vgrad}
[\nabla v_n(\bbw)]_i=\tfrac{1}{R_{ii}}\,\ell_{n,i}\bigl(\bbbeta_n^\star(\bbw)\bigr),
\end{equation}
the (noise-normalized) residual of the deployed estimator at measurement $i$. Replacing the LL optimality constraint by a penalized optimality gap yields the single-level problem over the design $\bbw$ \textit{and} the ``free'' reconstruction copies $\bbbeta=[\bbbeta_1,\ldots,\bbbeta_N]$:
\begin{equation}\label{eq:penalized}
\min_{\bbw\in[0,1]^M,\,\bbbeta}\ \!\!\! \Phi(\bbbeta) + \eta\sum_{i=1}^M\varphi(w_i)
+\gamma\sum_{n=1}^N\bigl(g_n(\bbw,\bbbeta_n)-v_n(\bbw)\bigr),
\end{equation}
where $(g_n(\bbw,\bbbeta_n)-v_n(\bbw))$ is the so-called optimality gap and $\gamma>0$ controls bilevel feasibility. Each gap term is nonnegative and vanishes if and only if $\bbbeta_n$ solves the LL problem. Moreover, for $\gamma$ large enough, solutions of \eqref{eq:penalized} are approximate solutions of \eqref{eq:sparse_blo} \cite{shen2023penalty,lu2024firstorder}. Crucially, $\Phi$ acts on the free copies, so no derivative of $\bbw\mapsto\bbbeta_n^\star(\bbw)$ is ever needed. The exact estimator enters only through $\nabla v_n$ in \eqref{eq:vgrad}, i.e., through residuals of (inexact) LL solves.

\section{Single-loop algorithm and guarantees}\label{sec:algorithm}
We set $\varphi=\varphi_\theta$ as in \eqref{eq:concave_price} and augment the objective of \eqref{eq:penalized} with the indicator function of the set $[0,1]^M$. The resulting function is then split into a smooth component
\[
\bar F(\bbw,\bbbeta):=\Phi(\bbbeta)+\eta\sum_i\varphi_\theta(w_i)+\gamma\sum_n\bigl(g_{{\rm s},n}(\bbw,\bbbeta_n)-v_n(\bbw)\bigr),
\]
and a nonsmooth, proximable component
\[
r(\bbw,\bbbeta):=\iota_{[0,1]^M}(\bbw)+\gamma\lambda\sum_n\|\bbbeta_n\|_1,
\]
with $\iota_{\ccalX}(\cdot)$ being the indicator function over set $\ccalX$. A proximal-gradient step on $(\bbw,\bbbeta)$ consists of a gradient step followed by entrywise soft-thresholding for each $\bbbeta_n$ and projection of $\bbw$ onto $[0,1]^M$ by coordinatewise clipping.
The gradients with respect to each $\bbbeta_n$ and to $\bbw$ are given by
\begin{align}
	&\nabla_{\bbbeta_n}\bar F(\bbw,\bbbeta)
	=\tfrac1N\bbX_{{\rm v},n}^\top(\bbX_{{\rm v},n}\bbbeta_n-\bby_{{\rm v},n})
	+\gamma\nabla_{\bbbeta_n}g_{{\rm s},n}(\bbw,\bbbeta_n),\nonumber \\
	&[\nabla_\bbw \bar F(\bbw,\bbbeta)]_i
	\!=\! \eta\,\varphi_\theta'(w_i)
	\!+\! \tfrac{\gamma}{R_{ii}}\sum_{n=1}^N\bigl(\ell_{n,i}(\bbbeta_n) \!-\ell_{n,i}(\hbbeta_n)\bigr), \label{eq:wgrad}
\end{align}
Here $\hbbeta_n\approx\bbbeta_n^\star(\bbw)$ is an inexact LL solution produced by $T_k$ warm-started forward-backward (FB) steps on the LL objective in \eqref{eq:ll}, which is the only place where the estimator is solved.

The price $\eta$ is applied as a \emph{homotopy} (starting with $\eta=0$ and reaching $\eta=\eta_J$, see details about $\eta$ in Alg.~\ref{alg:loop}).
The increasing schedule for $\gamma$ first runs price-free from the full design $\bbw=\bbone$. This way, the value-function coupling (the selection signal in \eqref{eq:wgrad}) is formed on well-fit reconstructions. The price level is then ramped geometrically at fixed $\gamma$, pruning the design progressively. Every ramp stage whose active set has settled contributes one deployable design, so a single run of Alg.~\ref{alg:loop} returns the whole cardinality vs.\ performance frontier.
An adaptive safeguard limits sensor removal at each price stage. If a price increment removes more than a fraction $\varrho$ of the active sensors, the stage is rolled back and repeated at the geometric mean of the previous and current price levels.
This prevents pruning avalanches from jumping across the informative segment of the frontier.

\begin{algorithm}[t]
\caption{Bilevel OED via a sparsity-price homotopy}\label{alg:loop}
\begin{algorithmic}[1]
\Require $\bbw=\bbone$, steps $\alpha_w,\alpha_\beta$, schedules $\{\gamma_t\}$, $\{\eta_j\}$ (geometric), inner schedule $\{T_k\}$, guard $\varrho$, threshold $\tau$
\For{$\gamma\in\{\gamma_1<\dots<\gamma_T\}$ with $\eta=0$, then $\eta\in\{\eta_1<\dots<\eta_J\}$ at $\gamma=\gamma_T$}
  \For{$k=1,\dots,K_{\rm stage}$}
  \State $\hbbeta_n \gets T_k$ warm-started FB steps on $g_n(\bbw,\cdot)$, $\forall n$
  \State $\bbbeta_n\!\gets\!\mathrm{soft}_{\alpha_\beta\gamma\lambda}\bigl(\bbbeta_n-\alpha_\beta\nabla_{\bbbeta_n}\bar F(\bbw\!,\bbbeta)\bigr)$, $\forall n$
  \State $\bbw\gets \mathrm{clip}_{[0,1]}\bigl(\bbw-\alpha_w\nabla_\bbw\bar F(\bbw,\bbbeta)\bigr)$ \Comment{uses \eqref{eq:wgrad}}
  \EndFor
  \If{price stage pruned $>\varrho\,|\!\supp(\bbw)|$ sensors}
  \State roll back; insert intermediate price $\sqrt{\eta_{j-1}\eta_j}$
  \Else{ record frontier design $\hbw=\bbone[\bbw>\tau]$}
  \EndIf
\EndFor
\end{algorithmic}
\end{algorithm}

\begin{remark}[Measurements must pay for themselves]\label{rem:price}
At any stationary point of \eqref{eq:penalized}, box optimality at $w_i=0$ requires $[\nabla_\bbw\bar F]_i\ge 0$, i.e., by \eqref{eq:wgrad}, measurement $i$ stays discarded exactly when
\begin{equation}\label{eq:price}
\frac{\gamma}{R_{ii}}\sum_{n=1}^N\Bigl(\ell_{n,i}\bigl(\hbbeta_n\bigr)-\ell_{n,i}\bigl(\bbbeta_n\bigr)\Bigr)\ \le\ \eta\,\varphi_{\theta}'(0)=\frac{\eta}{\theta} .
\end{equation}
The left-hand side is the aggregate amount by which the UL-corrected reconstructions $\bbbeta_n$ fit measurement $i$ better than the deployed estimator currently does. It can be read as an \emph{estimator-aware marginal value} of acquiring $i$, weighted by its noise level. A sensor is kept only if this value exceeds its marginal price. In addition,  a sensor at full weight is pushed down only by $\eta\varphi_\theta'(1)=\eta\theta/(1+\theta)^2\ll\eta/\theta$. Entering the design is thus much harder than staying in it, which keeps the selection stable along the price path.
Since $\Phi$ in \eqref{eq:phi} has no explicit $\bbw$-dependence, this coupling is the \emph{only} channel through which estimator performance influences the design, rendering the deployed estimator, not a surrogate model, central to determining which measurements matter.
\end{remark}

\noindent\textbf{Convergence guarantees.} For every iteration $k$, Alg.~\ref{alg:loop} requires $T_k$ FB steps to compute the approximate LL solution $\hbbeta_n$ [cf. line 3]. Next, we adapt the results in \cite{rey2026singlelevel} to characterize the solution generated by Alg.~\ref{alg:loop} after $K$ iterations. To be specific, let $\alpha=\min\{\alpha_w,\alpha_\beta\}$ and let $c>0$ be an arbitrary constant, which only affects the constant in the rate. We say that $(\bbw,\bbbeta)$ is $\varepsilon$-approximately bilevel feasible if $\sum_n[g_n(\bbw,\bbbeta_n)-v_n(\bbw)]\le\varepsilon$, i.e., if the reconstruction copies are $\varepsilon$-close, in LL cost, to the deployed estimates. 

\begin{proposition}[Convergence]\label{prop:rate}
Suppose that: a) each fidelity $g_{{\rm s},n}(\bbw,\cdot)$ is $\mu$-strongly convex and $L$-smooth on the region visited by the iterates; b) $\Phi$ and $\nabla_\bbw g_{{\rm s},n}$ are Lipschitz there; and c) at every iteration $k$, the inner-solve gap satisfies $\sum_n[g_n(\bbw^k,\hbbeta_n^k)-v_n(\bbw^k)]\le c/(\alpha\gamma k)^2$. Then, with appropriate constant steps, the proximal-gradient mapping $\bbG^k=\bbG(\bbw^k,\bbbeta^k)$ of \eqref{eq:penalized} satisfies $\frac1K\sum_{k=1}^K\|\bbG^k\|_2^2 \!=\! O(1/K)$. Moreover, for $\gamma\gtrsim 1/\sqrt{\mu\varepsilon}$, stationary points of \eqref{eq:penalized} are $\varepsilon$-approximate bilevel-feasible points of \eqref{eq:sparse_blo}.
\end{proposition}

\noindent Note that condition c) holds with $T_k=O(\log k)$ because the FB iteration contracts linearly under strong convexity. The proof specializes the inexact proximal-gradient analysis of \cite{rey2026singlelevel} (see also \cite{shen2023penalty,lu2024firstorder} for the penalty correspondence) and is omitted due to space constraints. The price term poses no difficulty: $\varphi_\theta$ has a Lipschitz gradient on the box, and the $\bbw$-prox is a coordinatewise clip rather than a capped-simplex projection. Each design on the price path is deployed as $\hbw=\bbone[\bbw>\tau]$ with $\tau=10^{-3}$. The box prox returns exact zeros, so $\tau$ only formalizes the support, while the surviving (possibly fractional) weights are set to
one, which is conservative at matched cardinality $M_1=\|\hbw\|_0$. All results in Sec.~\ref{sec:experiments} are reported for the deployed binary designs $\hbw$.

\section{Numerical experiments}\label{sec:experiments}

\begin{figure*}[t!]
	\centering
		\includegraphics[width=0.31\textwidth]{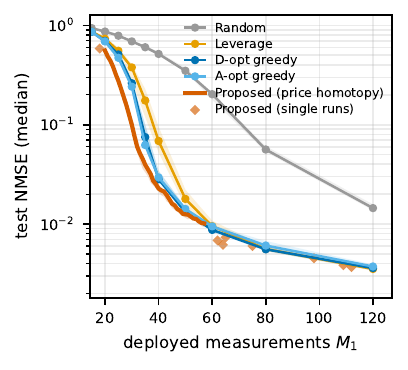}
	\hfill
		\includegraphics[width=0.31\textwidth]{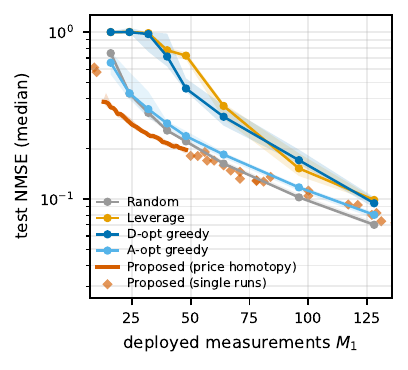}
	\hfill
		\includegraphics[width=0.31\textwidth, trim=0 -3.5mm 0 0]{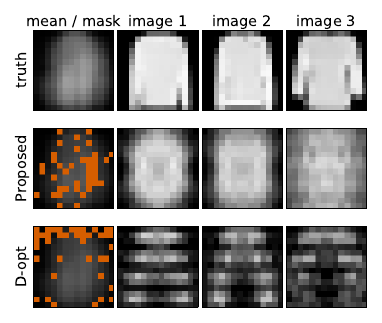}
    \vspace{-.2cm}
	\caption{(Left) Sparse linear test case: median test NMSE vs.\ $M_1$
		(bands: IQR over $10$ draws). Proposed frontier solid where at least 5
		draws contribute a design, diamonds otherwise; it starts at $M_1=240$
		(off scale). (Center) Fashion-MNIST: median test NMSE vs.\ $M_1$;
		conventions as in the left panel. (Right) Fashion-MNIST, $M_1=40$ (split 0):
		masks (orange, over the mean image) and reconstructions of 3 test images.}
	\label{fig:combined_results}
	\vspace{-.4cm}
\end{figure*}

This section studies two key aspects of the proposed approach.
First, it assesses the estimation accuracy of the selected designs relative to classical information-based methods at matched cardinality; and second, it evaluates the ability of the price homotopy to explore the trade-off between measurement cardinality and estimation accuracy.
All designs are evaluated using the same deployed estimator \eqref{eq:ll} on independent test instances, so that the comparisons isolate the effect of measurement selection.
We report medians and interquartile ranges (IQR) over $10$ random draws of the data.
Selected baselines include: random selection, leverage (row-norm) sampling, and greedy D- and A-optimal selection on $\bbF(\bbw)$, evaluated on a grid of cardinalities $M_1$ covering the frontier.
For the proposed method, “Proposed” aggregates errors from designs obtained in at least five different draws at a given cardinality, while “single draws” shows individual designs.
Code and full experimental details are available online.\footnote{{\scriptsize \url{https://github.com/reysam93/sparse_bilevel_oed}}}

\noindent\textbf{Sparse linear inverse problem.}
The first test case is a sparse linear regression with $M=240$ sensors (candidate measurements) and $D=80$ unknowns.
The rows and columns of $\bbX\in\reals^{240\times 80}$ are split into four groups, resulting in sensor families of $60$ rows, with each row comprising 4 blocks of 20 columns (unknowns).
The entries of $\bbX$ are zero-mean Gaussian where, for family $k$, the entries in the $k$-th block of columns have unit variance and the remaining entries have variance $1/25$.
Hence, each sensor family observes one block of unknowns well and the other three only weakly.
Signals $\bbbeta_n$ have $8$ nonzero entries, $70\%$ of which fall in two of the four blocks. Noise is white with a  $20$\,dB SNR. Each draw contains $N=64$ training, $64$ validation, and $512$ test instances, and we report medians over $10$ draws.  We set $\mu=0.1$ and $\lambda/\lambda_{\max}=0.1$, so that the deployed elastic net is genuinely nonsmooth.

The left panel in Fig.~\ref{fig:combined_results} plots the median test normalized mean-square error (NMSE) of every method against $M_1$ (shaded areas: IQR).
For every $M_1<40$, the proposed designs attain a lower NMSE than all baselines.
To quantify the gains, note that each homotopy run yields one binary design per price stage that changes $M_1$: $6$ to $10$ designs per draw, $80$ in total, with $M_1\in[22,239]$ ($67$ with $M_1\le40$). Each design is compared with the interpolated baseline curve of the same draw at the same $M_1$, and we report the median relative NMSE reduction over designs. For the $67$ designs with $M_1\le40$, the reduction is $41\%$ w.r.t.\ D-optimal and $43\%$ w.r.t.\ A-optimal selection (IQR $31$--$50\%$ and $36$--$49\%$; lower NMSE in $67$ of $67$ comparisons for both), $35\%$ w.r.t.\ leverage sampling ($64$ of $67$), and $51\%$ w.r.t.\ random selection ($67$ of $67$). For $M_1>40$ the gains shrink as all methods approach the full-design error (median $19\%$ w.r.t.\ D-optimal). The homotopy prunes quickly while the design is redundant ($M_1$ from $240$ to under $100$ in a few stages), then advances by one to three measurements per stage, covering $M_1\in[22,37]$ densely; one run per draw yields all designs, whereas each baseline curve requires one solve per cardinality.

\noindent\textbf{Pixel selection on Fashion-MNIST.}
The second study selects \emph{pixels} for image reconstruction on the public Fashion-MNIST dataset, downsampled to $14\times14$ ($D=M=196$). Images are represented on an orthonormal 2-D DCT basis, candidate measurements are individual pixels observed at $30$\,dB SNR, and the deployed estimator is the elastic net on the DCT coefficients ($\mu=10^{-2}$ and the same strong-Lasso level $\lambda/\lambda_{\max}=0.1$ as above). Designs are learned from $128$ training and $128$ validation images and evaluated on $512$ independent test images over $10$ splits. This setting is adversarial for model-based design: the sensing rows are orthonormal, so every pixel carries identical energy and information criteria can barely rank pixels, whereas the value of a pixel is dictated by the \emph{class structure} of the images, which is what the validation-based UL sees. Indeed, greedy D-optimal selection clusters its pixels in low-information border rows (Fig.~\ref{fig:combined_results}, right) and random sampling becomes the strongest classical baseline (Fig.~\ref{fig:combined_results}, center). The homotopy reaches $9$ to $17$ pixels in every split. Below $20-30$ pixels the deployed estimator returns an all-zero estimate on a few test images ($3$ of $512$ at $16$ pixels); the baselines collapse much earlier and more often (D-optimal: $96$ of $512$ images at $32$ pixels, $182$ at $16$). Table~\ref{tab:fashion} reports the paired comparison. Below $48$ pixels the proposed designs improve on every baseline in every case; at abundant budgets ($M_1>64$) they tie with random sampling and still improve on the information criteria by $10$ to $41\%$.
\begin{table}[t]\centering\footnotesize\setlength{\tabcolsep}{3pt}
\caption{Fashion-MNIST: proposed designs vs.\ baselines for different ranges of $M_1$. Percentages are the median paired reduction of the test NMSE w.r.t.\ each baseline. In parentheses, the number of proposed designs that attain a lower test NMSE than the baseline over the total number of proposed designs in that range (each design is compared with the baseline of the same split at the same $M_1$). 	\vspace{-.2cm}}\label{tab:fashion}
\begin{tabular}{lcccc}\toprule
 & $M_1\le32$ & $32<M_1\le48$ & $48<M_1\le64$ & $M_1>64$\\ \midrule
Random   & 36\% (49/49) & 19\% (23/23) & 8\% (6/6) & $0$\% (6/12)\\
Leverage & 71\% (49/49) & 73\% (23/23) & 63\% (6/6) & 41\% (12/12)\\
D-opt    & 70\% (49/49) & 70\% (23/23) & 49\% (6/6) & 40\% (12/12)\\
A-opt   & 37\% (49/49) & 25\% (23/23) & 14\% (6/6) & 10\% (9/12)\\
\bottomrule\end{tabular}\end{table}

\vspace{-.07cm}
\section{Conclusions}\label{sec:conclusions}
\vspace{-.12cm}
This paper addressed OED for nonsmooth estimators using a continuous bilevel-optimization approach. The LL computes the deployed estimator for a given design (the observations to sample), and the UL scores the resulting estimate and optimizes over designs. We developed a single-loop value-function penalty algorithm that i) never differentiates the nonsmooth solution map and ii) retains the $O(1/K)$ guarantee of BLO problems with smooth LL. We considered a budget-free formulation in which acquisition weights are charged a concave sparsity price, reducing sensor selection to an interpretable value-vs-price test. Moreover, one price homotopy traces the whole cardinality vs.\ performance frontier. Experiments on sparse recovery and public image data showed that the designs shaped by the deployed estimator outperform classical information-driven selection at matched cardinality. Budgeted designs, rounding guarantees, other estimator-aware criteria, and TV-type regularizers are developed in the journal version of this work.

\vfill\pagebreak
\section{Compliance with Ethical Standards}

The experiments in this paper are numerical simulation studies for which no ethical approval was required.

\bibliographystyle{IEEEbib}
\bibliography{refs}

\end{document}